# Calculation of light transmittance in a film: considerations of the coating geometry, the agent distribution, and its probability density distribution

Ken-ichi Amano[a*]

[a] *Faculty of Agriculture, Meijo University, Nagoya, Aichi 468-8502, Japan.*

* Correspondence author: K. Amano (amanok@meijo-u.ac.jp)

**ABSTRACT**

Transmittance is an important parameter for various films such as sunscreen films and creams, biofilms, coating materials, etc. Even if amounts of a sunscreen agent are the same, the transmittance greatly changes depending on the coating geometry (CG) and the agent distribution (AD) in the film. In this study, we calculate the transmittance considering CG and AD. In addition, we associate the transmittance with probability density distribution of the thickness of the film. We found analytical and numerical solutions of the transmittance in several model cases. It can be used for prediction of performance of the sunscreen film and for a fair comparative evaluation. Mathematical techniques in calculation of the transmittance are also explained in detail.

# 1. Introduction

Developments of sunscreen films and creams, biofilms, and coating materials are important for a better life for us. The parameter of transmittance is one of the necessary things for the development, because it is related to the performances of the products. The parameter is also significant in developments of solar cells and photocatalysts. However, the transmittance largely depends on the coating geometry (CG) and the agent's distribution (AD). Hence, it is important to understand effect of CG and AD on the transmittance. In this study, we calculate the transmittance considering CG, AD, and the probability density function (PDF) of the thickness of the film (agent). We found analytical and numerical solutions of the transmittance in several model cases. It can be used for prediction of performance of the sunscreen film and for a fair comparative evaluation. In Chapter 2, the calculation theory of the transmittance is explained. In Chapter 3, the relationship between the transmittance and CG (AD) is shown. The analysis result related to PDF of the thickness of the film (agent) is also shown here. In Chapter 4, conclusion of this study and some comments are written. In *Appendix A-C*, mathematical relationships derived in the process of this study are written.

# 2. Theory
## *2.2. Film on a flat surface*

To calculate the transmittance, we consider a following film shown in Fig. 1. In the model, the lower interface is completely flat, while the upper interface is modeled as a trigonometric function (sine or cosine function). Fig. 1 shows CG of the film, but it can be seen as AD within a flat film. The green solid line and $t_0$ both represent the average height of the film. The parameter $a$ is half amplitude of the trigonometric function and $t(x')$ represents the thickness as a function of position $x'$. For geometric reason, range of $a$ is expressed as $0 \leq a \leq t_0$ ($-t_0 \leq a \leq t_0$ is also possible but we define $a$ as the value that is greater than or equal to 0 for explanatory simplicity). The geometry is not necessary to follow the trigonometric function, but the thickness must follow the arcsine distribution in the theory explained here. The model indicates that amounts of the agent are the same if each $t_0$ is the same. It has been reported that the thickness of the sunscreen cream on a biological skin accords with inverse gamma distribution [1][2]. Thus, Fig. 1 is not the model of such a sunscreen cream, but that of an engineered film surface [3][4][5]. For simplicity, we use only a sine function as a trigonometric function.

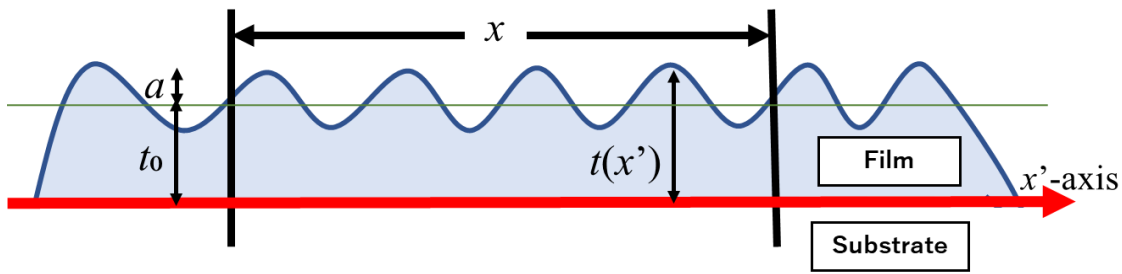

**Fig. 1.** CG of the film (it can also be seen as AD in a flat film). The upper coating interface is expressed as a trigonometric function (sine or cosine function) and that of the lower is flat. The green solid line and $t_0$ both represent the average height of the film.

As mentioned above, the geometry is not necessary to follow the trigonometric function, but the thickness must follow the arcsine distribution. It means that the upper interface does not need to be a periodic shape. If we disregard reflection and scattering from the film, the transmittance ($T$) can be calculated by using following equation:

$$T = \frac{1}{x}\int_0^x 10^{-\kappa t(x')}\,dx' = \frac{1}{x}\int_0^x e^{-\kappa t(x')\ln 10}\,dx', \quad (1)$$

where $\kappa$ (m$^{-1}$) is extinction coefficient (turbidity). The integration range is from 0 to $x$, but from 0 to $2\pi n$ ($n$ is a natural number) is also applicable. We are now considering the sufficiently long integration range ($0 \ll x$). The thickness function can be given by $t(x') = t_0 + a\sin(cx')$. For simplicity, $c$ is defined as a natural number, definition of which is not so unnatural because there are numerous humps within the beam cross section (within the integration range). Substituting above settings, Eq. (1) is rewritten as

$$\begin{aligned} T &= \frac{1}{2\pi n}\int_0^{2\pi n} e^{-\kappa(\ln 10)[t_0 + a\sin(cx')]}\,dx' \\ &= \frac{e^{-\kappa t_0 \ln 10}}{2\pi n}\int_0^{2\pi n} e^{-\kappa a(\ln 10)\sin(cx')}\,dx' \\ &= \frac{10^{-\kappa t_0}}{2\pi}\int_0^{2\pi} e^{-\kappa a(\ln 10)\sin(cx')}\,dx'. \quad (2) \end{aligned}$$

It is likely that Eq. (2) is related to an integral form of the Bessel function of the first kind of 0th order ($J_0$):

$$J_0(z) = \frac{1}{2\pi}\int_{-\pi}^{\pi} e^{iz\cos\theta}\,d\theta. \quad (3)$$

Since Eq. (3) can be rewritten as

$$J_0(bi) = \frac{1}{2\pi}\int_{-\pi}^{\pi} e^{-b\cos\theta}\,d\theta = \frac{1}{2\pi}\int_0^{2\pi} e^{-b\sin\theta}\,d\theta, \quad (4)$$

Eq. (2) is expressed as

$$T = 10^{-\kappa t_0} J_0(i\kappa a \ln 10). \qquad (5)$$

An interesting thing is that the parameter $c$ is disappeared in Eq. (5). That is, the transmittance does not depend on $c$ ($c$ was defined as a natural number in advance). Using Eq. (5), we can calculate the maximum value of the transmittance ($T_{\max}$). Substituting the maximum value of the half amplitude $a = t_0$, we obtain

$$T_{\max} = \lim_{a \to t_0} T(a) = 10^{-\kappa t_0} J_0(i\kappa t_0 \ln 10). \qquad (6)$$

On the other hand, the minimum value of the transmittance ($T_{\min}$) can be calculated when $a = 0$: $T_{\min} = 10^{-\kappa t_0}$. For instance, when $\kappa = 1.7 \times 10^5$ (m$^{-1}$) and $t_0 = 10^{-5}$ (m), $T_{\max}/T_{\min} \approx 10.5$. Therefore, we can understand that geometrical control is important for control of the transmittance parameter. It should be noted that this conclusion is not only for the sine function but also for a cosine function and a thickness function that obeys the arcsine distribution.

### 2.2. Film on a non-flat surface

We consider a following film model (see Fig. 2). In the model, both of the interfaces are not flat but waving. Fig. 2 shows CG of the film, but it can be seen as AD within a flat film. The parameters $a_1$ and $a_2$ are half amplitudes of the trigonometric functions. The green solid lines indicate the average heights of the upper and the lower interfaces, the interval between them is $t_0$. In this case, the transmittance is expressed as

$$T = \frac{1}{x}\int_0^x e^{-\kappa(\ln 10)t(x')}\,dx' = \frac{1}{x}\int_0^x e^{-\kappa(\ln 10)[t_1(x')-t_2(x')]}\,dx', \quad (7)$$

where $t_1(x') = t_{01} + a_1\sin(c_1 x')$ and $t_2(x') = -(t_0 - t_{01}) \mp a_2*\sin(c_2 x')$. The parameter ranges of $a_1$ and $a_2$ are $0 \leq a_1 \leq t_{01}$ and $0 \leq a_2 \leq t_0 - t_1$, respectively. Also in this case, we define $c_1$ and $c_2$ as natural numbers. It is not so unnatural because there are numerous humps within the beam cross section (within the integration range). To solve Eq. (7), we change it as follows:

$$\begin{aligned}
T &= \frac{1}{2\pi n}\int_0^{2\pi n} e^{-\kappa(\ln 10)[t_0 + a_1\sin(c_1 x') \pm a_2\sin(c_2 x')]}\,dx' \\
&= \frac{e^{-\kappa t_0 \ln 10}}{2\pi n}\int_0^{2\pi n} e^{-\kappa a_1(\ln 10)\sin(c_1 x')}\cdot e^{-\kappa a_2(\ln 10)\sin(c_2 x')}\,dx' \\
&= \frac{10^{-\kappa t_0}}{2\pi}\int_0^{2\pi} e^{-\kappa(\ln 10)[a_1\sin(c_1 x') \pm a_2\sin(c_2 x')]}\,dx'. \quad (8)
\end{aligned}$$

However, it is considered that Eq. (8) cannot be solved *analytically* any further (it can be solved numerically). To obtain the analytical solution, we introduce a following restriction: $c_1 \gg c_2$. In this case, the thickness of the film in Fig. 2 can be redrawn as Fig. 3. Since each length of the red line ($\Delta x'$) is sufficiently short and each blue curve mounting on the red line obeys the arcsine distribution, the partial transmittance at $x'$ in the range from $x'$ to $x' + \Delta x'$ is calculated as

$$\begin{aligned}
\frac{\Delta T(x')}{\Delta x'} &= \frac{1}{\Delta x'}\int_{x'}^{x'+\Delta x'} 10^{-\kappa[t_0 \pm a_2\sin(c_2 x') + a_1\sin(c_1 x'')]}\,dx'' \\
&= \frac{1}{2\pi/c_1}\int_{x'}^{x'+2\pi/c_1} 10^{-\kappa[t_0 \pm a_2\sin(c_2 x') + a_1\sin(c_1 x'')]}\,dx'' \\
&= \frac{1}{2\pi/c_1} 10^{-\kappa[t_0 \pm a_2\sin(c_2 x')]}\int_0^{2\pi/c_1} 10^{-\kappa[a_1\sin(c_1 x'')]}\,dx'' \\
&= \frac{1}{2\pi} 10^{-\kappa[t_0 \pm a_2\sin(c_2 x')]}\int_0^{2\pi} 10^{-\kappa[a_1\sin(c_1 x'')]}\,dx''
\end{aligned}$$

$$= 10^{-\kappa[t_0 \pm a_2 \sin(c_2 x')]} J_0(i\kappa a_1 \ln 10). \quad (9)$$

Hence, the total transmittance is calculated as

$$T = \frac{1}{2\pi n}\int_0^{2\pi n} \frac{\Delta T(x')}{\Delta x'} dx'$$
$$= \frac{1}{2\pi n}\int_0^{2\pi n} 10^{-\kappa[t_0 \pm a_2 \sin(c_2 x')]} J_0(i\kappa a_1 \ln 10)\, dx'. \quad (10)$$

It can be rewritten as

$$T = \frac{1}{2\pi n} J_0(i\kappa a_1 \ln 10)\int_0^{2\pi n} e^{-\kappa(\ln 10)[t_0 \pm a_2 \sin(c_2 x')]} dx'$$
$$= \frac{1}{2\pi} J_0(i\kappa a_1 \ln 10)10^{-\kappa t_0}\int_0^{2\pi} e^{\mp \kappa a_2(\ln 10)\sin(c_2 x')} dx'$$
$$= 10^{-\kappa t_0} J_0(i\kappa a_1 \ln 10) J_0(\pm i\kappa a_2 \ln 10). \quad (11)$$

Consequently, the analytical solution, the purpose of this section, is obtained. We note that "±" in $J_0$ can be omitted, because it is an even function. As shown in Eq. (11), the single integral in Eq. (7) was split into two integrals. In *Appendix A*, the split method in the integral is more explained.

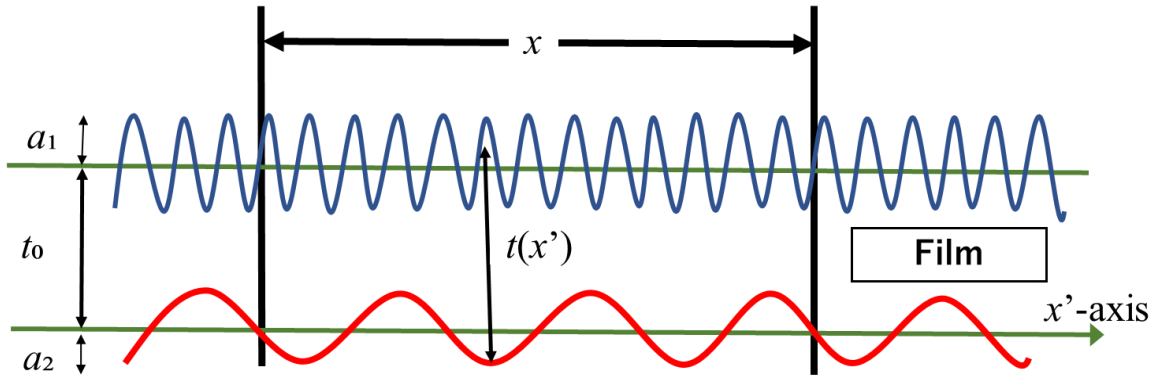

**Fig. 2.** CG of the film (it can also be seen as AD in a flat film). The upper and the lower coating

interfaces are expressed as trigonometric functions (sine or cosine functions).

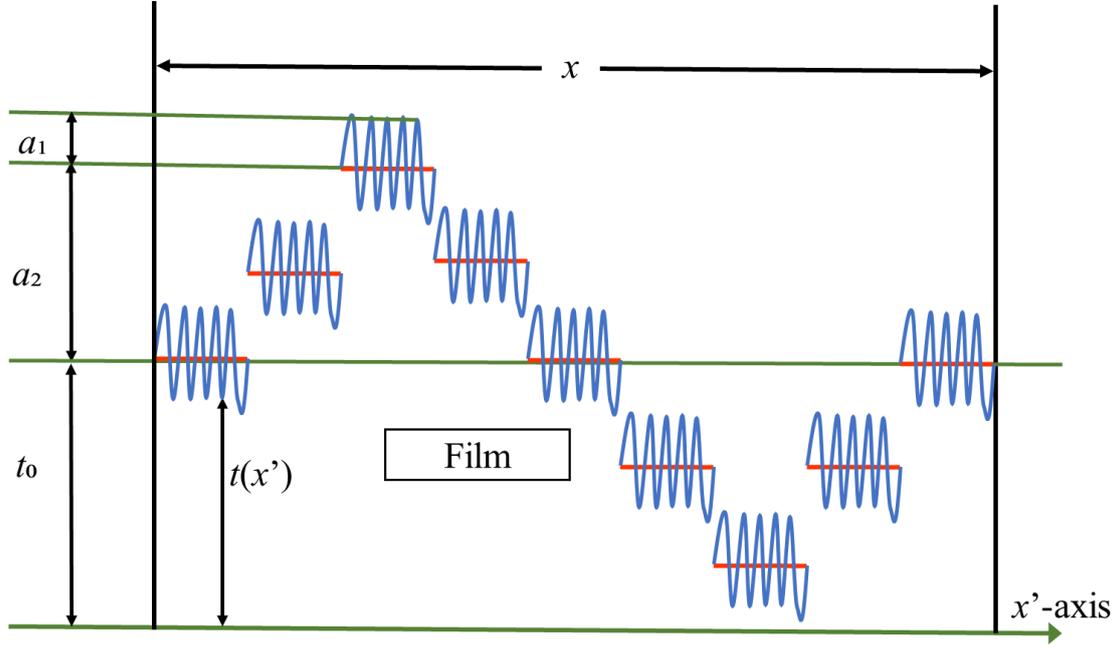

**Fig. 3.** Geometrically transformed film thickness. The thickness distributions of Figs. 2 and 3 are the same when $c_1 \gg c_2$. Colors of blue and red are related to those in Fig. 2. Each length of the red line ($\Delta x'$) is sufficiently short. The blue curve is mounting on the red short line, height distribution of which obeys the arcsine distribution.

By the way, we replace $T$ in Eq. (11) with $T_W$, and then we obtain

$$T_W = 10^{-\kappa t_0} J_0(i\kappa a_1 \ln 10) J_0(i\kappa a_2 \ln 10). \quad (12)$$

The subscript W represents that both of the upper and the lower interfaces are waving. Next, we prepare a film that has the same half amplitude $a_1$ as $T_W$ but $a_2$ is zero (the lower interface is flat). The transmittance of such a film ($T_S$) is given by

$$T_S = 10^{-\kappa t_0} J_0(i\kappa a_1 \ln 10), \quad (13)$$

where the subscript S represents that the only one interface is waving. Then, ratio of $T_W/T_S$ is calculated as

$$T_W/T_S = J_0(i\kappa a_2 \ln 10). \quad (14)$$

The ratio of $T_W/T_S$ can be used for prediction of the transmittance of a film that has the waves on both interfaces from the transmittance of a film that has the waving and flat surfaces on respective interfaces, and vice versa. Moreover, if we use ratio of $T_{min}/T_W$, the transmittance of a film that has flat interfaces on both side can be predicted from the transmittance of a film that has the waves on both sides. These ratios are important for a fair comparative evaluation of the product performance.

We note that the way of thinking shown in Fig. 3 and Eq. (12) can be applied to speed up of a numerical integration. If the integrand possesses a property shown in Fig. 3, the numerical integration can be divided into two simple parts. The more explanation is written in *Appendix A*.

### 2.3. PDF of the thickness

PDF of the thickness ($p(t)$) is important information in terms of CG. As the name implies, $p(t)$ takes following property:

$$\int_{t_{min}}^{t_{max}} p(t) dt = 1, \quad (15)$$

where $t_{min}$ and $t_{max}$ represent the minimum and the maximum values in $t$, respectively. The thickness takes positive values, and hence the integration range from 0 to $\infty$ is also applicable in Eq. (15). The transmittance can be also calculated by using $p(t)$, which is given below (the derivation process is explained in *Appendix B*):

$$T = \int_{t_{min}}^{t_{max}} e^{-\kappa(\ln 10)t} p(t) dt. \quad (16)$$

It is known that PDF of a sine function is an arcsine distribution. Thus, the transmittance of the film model shown in Fig. 1 can be calculated by substituting the arcsine distribution into $p(t)$, where $t_{min}$ and $t_{max}$ are $t_0 - a$ and $t_0 + a$, respectively.

There is the other calculation way of the transmittance, which is expressed by using an inverse function of cumulative PDF ($b(u)$) (the derivation process is explained in *Appendix B*):

$$T = \int_0^1 e^{-\kappa(\ln 10)b(u)} du. \quad (17)$$

$b(u)$ is called the thickness as a function of the cumulative density function (CDF). An important thing is that there are three ways to calculate the transmittance: Eq. (1), Eq. (16), and Eq. (17).

As mentioned above, the transmittance of the film shown in Fig. 1 can be calculated by substituting the arcsine distribution into $p(t)$. However, can we obtain $p(t)$ when the film model is Fig. 2? The answer is yes. $p(t)$ can be calculated by following a process below: (I) Numerically or analytically calculate the transmittance by using Eq. (7), the solution of which is a function of "$\kappa \ln 10$"; (II) View Eq. (16) as the Laplace transform,

$$T = \int_0^\infty e^{-(\kappa \ln 10)t} p(t) dt, \quad (18)$$

which can be realized due to the domain of definition and its range of $p(t)$; (III)

Perform the inverse Laplace transform numerically or analytically, and then $p(t)$ can be obtained.

Acquisition process of PDF in the case of $c_1 \gg c_2$ is also explained. In this case, the transmittance can be readily obtained from Eq. (12), which can be rewritten as

$$T_W = 10^{-\kappa t_0} J_0(i\kappa a_1 \ln 10) J_0(i\kappa a_2 \ln 10)$$

$$= e^{-\kappa t_0 \ln 10} \frac{1}{2\pi} \int_0^{2\pi} e^{-\kappa a_1 (\ln 10) \sin\theta} d\theta \frac{1}{2\pi} \int_0^{2\pi} e^{-\kappa a_2 (\ln 10) \sin\theta} d\theta$$

$$= e^{-\kappa (a_1 + t_0 - a_1) \ln 10} \frac{1}{2\pi} \int_0^{2\pi} e^{-\kappa a_1 (\ln 10) \sin\theta} d\theta \frac{1}{2\pi} \int_0^{2\pi} e^{-\kappa a_2 (\ln 10) \sin\theta} d\theta$$

$$= \frac{1}{2\pi} \int_0^{2\pi} e^{-\kappa (\ln 10) a_1 (1+\sin\theta)} d\theta \frac{1}{2\pi} \int_0^{2\pi} e^{-\kappa (\ln 10) a_2 \{[(t_0 - a_1)/a_2] + \sin\theta\}} d\theta, \quad (19)$$

where Eq. (4) is used for the derivation. Eq. (19) can be replaced using Eq. (16) as

$$T_W = \frac{1}{2\pi} \int_0^{2\pi} e^{-\kappa (\ln 10) a_1 (1+\sin\theta)} d\theta \frac{1}{2\pi} \int_0^{2\pi} e^{-\kappa (\ln 10) a_2 \{[(t_0 - a_1)/a_2] + \sin\theta\}} d\theta$$

$$= \int_0^{2a_1} e^{-\kappa (\ln 10) t} p_1(t) dt \int_{t_0 - a_1 - a_2}^{t_0 - a_1 + a_2} e^{-\kappa (\ln 10) t} p_2(t) dt, \quad (20)$$

where $p_1(t)$ and $p_2(t)$ are PDFs of heights of "$a_1(1+\sin\theta)$" and "$a_2\{[(t_0 - a_1)/a_2] + \sin\theta\}$", respectively. The two integrals above can be seen as the Laplace transforms,

$$T_W = \int_0^\infty e^{-\kappa (\ln 10) t} p_1(t) dt \int_0^\infty e^{-\kappa (\ln 10) t} p_2(t) dt. \quad (21)$$

Mathematical form of Eq. (21) is related to the convolution integral in the Laplace transform. Hence, the inverse Laplace transform of $T_W$ as a function of "$\kappa \ln 10$" is given by

$$\frac{1}{2\pi i}\int_{\gamma-i\infty}^{\gamma+i\infty} e^{\kappa(\ln 10)t} T_W(\kappa\ln 10) d(\kappa\ln 10) = \int_0^t p_1(t')p_2(t-t')dt' = p_W(t). \quad (22)$$

Since the solution of the inverse Laplace transform of the transmittance is PDF of the film thickness (see Eq. (18)), the integral on the right-hand side of Eq. (22) yields PDF in the case of Fig. 3 ($p_W(t)$). From Eq. (22), a mathematical property that the convolution integral of two PDFs generates a new PDF is found. That is,

$$\int_0^\infty \int_0^t p_1(t')p_2(t-t')dt'dt = \int_0^\infty p_W(t)dt = 1. \quad (23)$$

The mathematical property above is one of the universal facts of PDF.

### 2.4. Thickness function with beat and its transmittance

Here, we challenge calculation of the transmittance in the case of $c_1 \approx c_2$ ($c_1 \neq c_2$) by applying the PDF integration. In Chapters 2.1, 2.2, and 2.3, we defined both $c_1$ and $c_2$ as natural numbers for simplicity and strictness. However, in this chapter, such restriction is eliminated. The elimination can be realized without problem, because the sufficiently long integration range has been introduced in the original equation. When $c_1 \approx c_2$ ($c_1 \neq c_2$), sum of the sine functions makes beat. For instance, the shapes of "$2\sin(2\pi x) + \sin(0.99 \times 2\pi x)$" and "$2\cos(2\pi x) + \cos(0.01 \times 2\pi x)$" resemble each other in some degree, where the former and the latter functions respectively represent the cases of $c_1 \approx c_2$ ($c_1 \neq c_2$) and $c_1 \gg c_2$. Their PDFs of the film thicknesses are greatly similar in visual. Hence, the transmittance in the case of $c_1 \approx c_2$ ($c_1 \neq c_2$) can be calculated by substituting PDF in the case of $c_1 \gg c_2$ into Eq. (16). Since Eq. (16) connects to Eq. (7), the transmittance in the case of $c_1 \approx c_2$ ($c_1 \neq c_2$) can be approximately calculated by using Eq. (12) derived in the case of $c_1 \gg c_2$. Using the PDF approximation, we were able to accurately calculate the transmittance up to 2 or 3 significant digits. The high accuracy proves validity of the PDF approximation, and it means that approximate shape of PDF of beat can be calculated by using Eq. (22) of the convolution integral. This may be great advantage for other scientific calculations containing the beat functions.

*2.5. Transmittances calculation with a pair of PDF models*

Until now, we have used the sine function and its PDF (arcsine distribution) as a model of the film. In this chapter, we additionally introduce the inverse gamma distribution, the gamma distribution, the exponential distribution, the beta distribution, the chi-squared distribution, the Weibull distribution [6], the Flory-Schulz distribution [7][8], etc. The inverse gamma distribution is known as a feasible model of the thickness distribution of the sunscreen cream on a biological skin [1][2]. If the lower interface is flat and PDF of the upper interface can be modeled as one of the PDFs mentioned above, the transmittance can be calculated by

$$T = \int_{t_{\min}}^{t_{\max}} e^{-\kappa(\ln 10)t} p_j(t)dt = \int_0^\infty e^{-(\kappa \ln 10)t} p_j(t)dt, \qquad (24)$$

where $p_j(t)$ represents the selected PDF and $0 \le t_{\min} < t_{\min}$. On the other hand, when the lower interface is also not flat, the transmittance can be calculated by

$$T = \int_{t_{\min}}^{t_{\max}} e^{-\kappa(\ln 10)t} p_{jk}(t)dt = \int_0^\infty e^{-(\kappa \ln 10)t} p_{jk}(t)dt, \qquad (25)$$

$$p_{jk}(t) = \int_0^t p_k(t')p_j(t-t')dt', \qquad (26)$$

where $p_k(t)$ and $p_{jk}(t)$ represent the selected PDF of the lower interface and the mixed

PDF. This calculation method can be utilized when geometrical relationship between the upper and the lower interfaces corresponds to the situation in Chapters 2.2 (Fig. 3) and II.D (beat). That is, when the oscillation lengths (cycle lengths) of the upper and the lower interfaces are significantly different or significantly close, the above two equations can be used for calculation of the transmittance.

## 3. Results and discussion

In this chapter, we show calculation results. Fig. 4 shows the ratio of transmittance vs amplitude of the upper sine function. In the red and green curves, the half amplitudes ($a_2$) are 0 μm and 5 μm, respectively. In all the cases, the average thicknesses are 10 μm. That is, the averaged amounts of the film material on a unit area are $10^{-3}$ cm$^3$/cm$^2$. The extinction coefficient $\kappa$ (m$^{-1}$) is set to $2\times10^5$ m$^{-1}$. As shown in the figure, the transmittance increases as the half amplitude increases. Furthermore, the slope of the green curve is steeper than that of the red curve. It means that the rougher the surfaces, the higher the transmittance. The result is caused by partially formed thin areas of the film. As a comparison, a numerically integrated result (black dotted curve) is also shown, where the analytical solution of Eq. (12) is not used. In the green curve, $c_1 \gg c_2$, while in the black dotted curve, $c_1/5 = c_2$. From the comparison between the green curve and black dotted curve, it can be said that if $c_1$ is about 5 times larger than $c_2$, the accuracy of the analytical solution (Eq. (12)) is quite high. This conclusion is confirmed from several conditions (not shown).

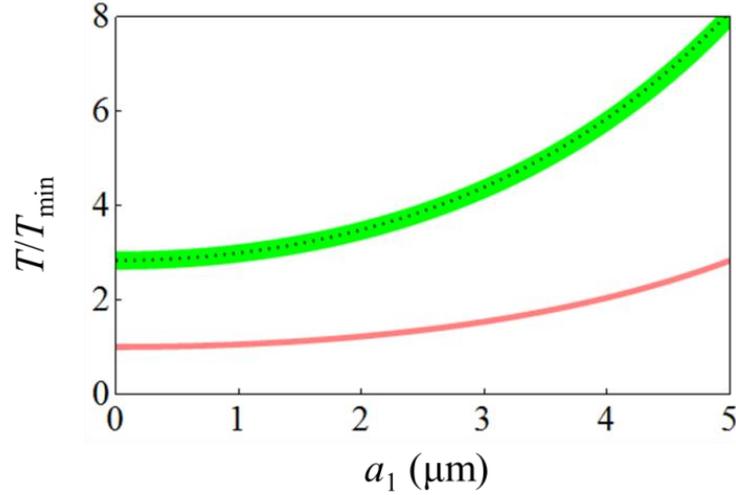

**Fig. 4.** Ratio of transmittance ($T/T_{min}$) vs half amplitude of the upper sine function $a_1$.

Next, we show PDF of the thickness calculated from Eq. (22). In addition, we check the shape of PDF in the case that both the upper and the lower interfaces are waving. The red curve ($p_W(t)$) in Fig. 5 represents PDF of the film thickness inversely calculated from a film that has waving interfaces on both sides. $p_W(t)$ is calculated from the thickness function $t(x') = 4 + \sin(c_1 x') + 2\sin(c_2 x')$ and $c_1 \gg c_2$. It is related to Fig. 2 with the condition $c_1 \gg c_2$, leading to Fig. 3. Arcsine distribution (blue curve) from 2 to 6 is also shown as a reference, where the thickness function is $t(x') = 4 + 2\sin(c_1 x')$. It is related to PDF of the film thickness in Fig. 1. PDF of the thickness in the case of the both upper and lower interfaces have waves decays like a mountain slope on both sides and has two peaks in the middle area. This shape is interesting compared with that that of the blue curve, because the shape is relatively unfamiliar. The shape of $p_W(t)$ changes depending on the substituted pair of PDFs.

As explained in Chapter 2.4, the convolution integral (Eq. (22)) can be used for the PDF generator for the situation of the beat. Hence, one can see that the red curve as the approximate form of PDF of the beat.

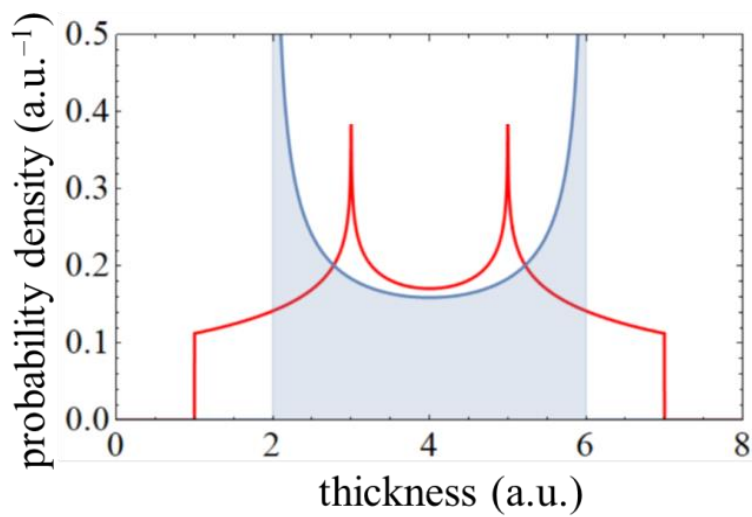

**Fig. 5.** $p_W(t)$ (red curve) calculated from the convolution integral (Eq. (22)) and arcsine distribution (blue curve) from 2 to 6.

## 4. Conculsion

In summary, we have shown calculation method of the transmittance by introducing several film models. It has been shown that the transmittance is increased as the roughness of CG (AD) is increased. The analysis result of PDF of the thickness has also been shown by using the inverse calculation method. The transmittance is one of the important parameters for developments of sunscreen films and creams, biofilms, coating materials, solar cells, and photocatalysts. The achievements obtained in this study are significant for a fair comparative evaluation of the products. Several mathematical techniques have been explained, which may be useful for other scientific calculations. In the future, we will use the mathematical techniques to challenge and solve several inverse problems in physical chemistry and biophysics.

**Acknowledgements**

We would like to thank M. Maebayashi for useful discussion and thank K. Sakai for helping preparation of Figs. 1-3. This work was supported by Grant-in-Aid for Academic Research from Meijo university and Grant-in-Aid for Scientific Research (C) from Japan Society for the Promotion of Science (No. 20K05437).

**Appendix**

Here, we show some mathematical relationships derived through the process of the present study. These relationships may be useful in other scientific calculations.

*Appendix A. Splitting of a single integral into two integrals*

We consider a following integral ($F_0(s)$),

$$F_0(s) = \frac{1}{x_2 - x_1} \int_{x_1}^{x_2} e^{-s[h_0 + h_1(x) + h_2(x)]} dx, \quad (27)$$

where $s$, $h_0$, $h_1(x)$, and $h_2(x)$ are a variable of $F_0$, a constant, arbitrary periodic functions 1 and 2, respectively. The integral is performed in following condition: $0 \leq x_1 \leq x_2$. If length of one cycle of the periodic function 1 ($\lambda_1$) is much shorter than that of the periodic function 2 ($\lambda_2$), the integral can be seen as Fig. 3. Then, the derivative function at $x$ in the range from $x$ to $x + \Delta x$ is calculated as

$$\frac{\Delta F_0(s; x)}{\Delta x} = \frac{1}{\Delta x} \int_{x}^{x+\Delta x} e^{-s[h_0 + h_1(x) + h_2(x)]} dx. \quad (28)$$

Since $\lambda_1$ is sufficiently shorter than $\lambda_2$, Eq. (28) can be rewritten as

$$\frac{\Delta F_0(s;x)}{\Delta x} = e^{-s[h_0+h_2(x)]} \frac{1}{\Delta x} \int_{x}^{x+\Delta x} e^{-sh_1(x)} dx. \quad (29)$$

When $\lambda_1$ is sufficiently shorter than $\Delta x$, $\Delta x$ can be expressed as $\Delta x = (n + \Delta n)\lambda_1$, where $n$ is a natural number and $0 \leq \Delta n \leq 1$. Then Eq. (29) becomes

$$\frac{\Delta F_0(s;x)}{\Delta x} = e^{-s[h_0+h_2(x)]} \frac{1}{(n+\Delta n)\lambda_1} \int_{x}^{x+(n+\Delta n)\lambda_1} e^{-sh_1(x)} dx$$

$$= e^{-s[h_0+h_2(x)]} \frac{1}{\lambda_1} \int_{0}^{\lambda_1} e^{-sh_1(x')} dx', \quad (30)$$

where we used approximation that $(n + \Delta n) = n$. This is because, $\lambda_1$ is sufficiently shorter than $\Delta x$, corresponding to that $n$ is sufficiently larger than $\Delta n$. Using Eq. (30), Eq. (27) can be rewritten as

$$F_0(s) = \frac{1}{x_2 - x_1} \int_{x_1}^{x_2} \frac{\Delta F_0(s;x)}{\Delta x} dx$$

$$= \frac{1}{x_2 - x_1} \int_{x_1}^{x_2} e^{-s[h_0+h_2(x)]} \frac{1}{\lambda_1} \int_{0}^{\lambda_1} e^{-sh_1(x')} dx' dx$$

$$= e^{-sh_0} \left( \frac{1}{x_2 - x_1} \int_{x_1}^{x_2} e^{-sh_2(x)} dx \right) \left( \frac{1}{\lambda_1} \int_{0}^{\lambda_1} e^{-sh_1(x')} dx' \right). \quad (31)$$

Therefore, the single integral can be split into two integrals when the situation of Fig. 3 ($\lambda_2 \gg \lambda_1$) is realized. In the present study, use of Eq. (31) enabled us to obtain the analytical solution of the transmittance (Eq. (11)). This mathematical technique is also applicable when sum of $h_1(x)$ and $h_2(x)$ generates beat ($\lambda_2 \approx \lambda_1$ and $\lambda_2 \neq \lambda_1$) and the integration range is sufficiently long (see Chapter 2.4).

Next, we explain a more general splitting method for an integral ($V$):

$$V = \int_{x_1}^{x_2} f_S(x) f_P(x)\, dx, \qquad (32)$$

where, $f_S(x)$ and $f_P(x)$ are an arbitrary smooth function and a periodic function, respectively. If $\Delta x$ is small enough, Eq. (32) can be rewritten as

$$V = \int_{x_1}^{x_2} \frac{\Delta V(x)}{\Delta x} dx. \qquad (33)$$

Assuming that $f_P(x)$ has short wave length $\lambda_P$ ($\equiv \Delta x$), and $f_S(x)$ is smooth enough to satisfy a condition that it is almost constant within the span $\lambda_P$, the derivative term can be expressed as

$$\begin{aligned}\frac{\Delta V(x)}{\Delta x} &= \frac{f_S(x)}{\Delta x} \int_x^{x+\Delta x} f_P(x')dx' \\ &= \frac{f_S(x)}{\lambda_P} \int_x^{x+\lambda_P} f_P(x')dx' = \frac{f_S(x)}{\lambda_P} \int_0^{\lambda_P} f_P(x')dx'. \end{aligned} \qquad (34)$$

Substituting Eq. (34) into Eq. (33), the integral in Eq. (32) can be divided into two integrals:

$$\int_{x_1}^{x_2} f_S(x) f_P(x)\, dx = \left[ \int_{x_1}^{x_2} f_S(x)\, dx \right] \left[ \frac{1}{\lambda_P} \int_0^{\lambda_P} f_P(x)dx \right]. \qquad (35)$$

For example, if $f_P(x)$ is white noise, trigonometric function, triangle wave, square wave, or sawtooth wave, the second integral in the right-hand side of Eq. (35) is zero. It means that approximated value of $V$ is also zero. On the other hand, if $f_P(x)$ is not such a symmetrical wave, the second integral is not zero. Then the approximated value of $V$ becomes non-zero value. We tested Eq. (35) numerically, and obtained an applicable result when $\lambda_P$ is sufficiently shorter than $(x_2 - x_1)$ of the integral range.

In the final paragraph, we propose an approximate form of Eq. (32) in a condition that $f_S(x)$ is not constant within the span $\lambda_P$. Here, we hypothecate that $f_P(x)$ obeys the following equation:

$$\int_x^{x+\lambda_P} f_P(x)\,dx = 0. \qquad (36)$$

For example, when $f_P(x)$ is trigonometric function, triangle wave, square wave, or sawtooth wave, Eq. (36) is satisfied. If $f_S(x)$ is smooth enough to satisfy a condition that it is almost a linear function within the span $\lambda_P$, the $\Delta V(x)/\Delta x$ in Eq. (33) can be expressed as

$$\begin{aligned}\frac{\Delta V(x)}{\Delta x} &= \frac{1}{\lambda_P}\int_x^{x+\lambda_P}\left[f_S(x) + \frac{df_S(x)}{dx}x'\right]f_P(x')dx' \\ &= \frac{f_S(x)}{\lambda_P}\int_x^{x+\lambda_P} f_P(x')dx' + \frac{df_S(x)}{dx}\frac{1}{\lambda_P}\int_x^{x+\lambda_P} x' f_P(x')dx' \\ &= \frac{df_S(x)}{dx}\frac{1}{\lambda_P}\int_x^{x+\lambda_P} x' f_P(x')dx'. \qquad (37)\end{aligned}$$

The integral of $V$ starts from $x_1$, and hence Eq. (37) can be approximately rewritten as

$$\frac{\Delta V(x)}{\Delta x} = \frac{df_S(x)}{dx}\frac{1}{\lambda_P}\int_{x_1}^{x_1+\lambda_P} x' f_P(x')dx'. \qquad (38)$$

Since $f_P(x)$ is the periodic function with the wave length $\lambda_P$, the following equation is satisfied: $f_P(x_1) = f_P(x_1 + \lambda_P)$. Moreover, if $(x_2 - x_1)/\lambda_P$ is an integer, the following equation also holds: $f_P(x_1) = f_P(x_2)$. Substituting Eq. (37) into Eq. (33), $V$ is rewritten as

$$V = \int_{x_1}^{x_2} \left[ \frac{df_S(x)}{dx} \frac{1}{\lambda_P} \int_{x_1}^{x_1+\lambda_P} x' f_P(x') dx' \right] dx$$

$$= \left[ \int_{x_1}^{x_2} \frac{df_S(x)}{dx} dx \right] \left[ \frac{1}{\lambda_P} \int_{x_1}^{x_1+\lambda_P} x' f_P(x') dx' \right]$$

$$= [f_S(x_2) - f_S(x_1)] \left[ \frac{1}{\lambda_P} \int_{x_1}^{x_1+\lambda_P} x' f_P(x') dx' \right]. \quad (39)$$

Eq. (38) is an approximate form of $V$. It indicates that if $f_S(x_1) = f_S(x_2)$, the approximate value of $V$ is zero. This form is helpful when integration of $f_S(x)f_P(x)$ is difficult to solve analytically. We tested Eq. (38) numerically, and obtained an applicable result when $\lambda_P$ is sufficiently shorter than $(x_2 - x_1)$ and $f_S(x)$ is sufficiently smooth. For reference, we also show an approximate form under the condition that integral range is from $x_1$ to $x_2 + \varphi$ ($|\varphi|$ is smaller than $\lambda_P$).

$$V = \int_{x_1}^{x_2+\varphi} f_S(x) f_P(x) \, dx$$

$$= [f_S(x_2) - f_S(x_1)] \left[ \frac{1}{\lambda_P} \int_{x_1}^{x_1+\lambda_P} x' f_P(x') dx' \right]$$

$$+ [f_S(x_2 + \varphi) - f_S(x_2)] \left[ \frac{1}{\varphi} \int_{x_2}^{x_2+\varphi} x' f_P(x') dx' \right]$$

$$+ \left[ \int_{x_2}^{x_2+\varphi} f_S(x) \, dx \right] \left[ \frac{1}{\varphi} \int_{x_2}^{x_2+\varphi} f_P(x') dx' \right]. \quad (40)$$

In the derivation process of Eq. (40), $f_S(x)$ in the range from $x_2$ to $x_2 + \varphi$ was also approximated as the linear function, but we found in many numerical calculation tests that it only should be approximated as the constant. In the numerical verification test of Eq. (40), we found that $f_S(x_2 + \varphi) - f_S(x_2)$ should be set to 0 (the second term should be deleted). Thus, the following approximated form is rather recommended when the integral range is from $x_1$ to $x_2 + \varphi$:

$$V = \int_{x_1}^{x_2+\varphi} f_S(x) f_P(x)\, dx$$
$$= [f_S(x_2) - f_S(x_1)] \left[ \frac{1}{\lambda_P} \int_{x_1}^{x_1+\lambda_P} x' f_P(x')\, dx' \right]$$
$$+ \left[ \int_{x_2}^{x_2+\varphi} f_S(x)\, dx \right] \left[ \frac{1}{\varphi} \int_{x_2}^{x_2+\varphi} f_P(x')\, dx' \right]. \quad (41)$$

The approximate form of the integral will be useful, because such kind of integrals are performed in many research areas (e.g., the integrals containing electromagnetic wave, acoustic wave, Bloch function, etc).

*Appendix B: Roles of PDF and an inverse function*

Here, we introduce an integral equation ($F_1(s)$),

$$F_1(s) = \frac{1}{x_2 - x_1} \int_{x_1}^{x_2} f(s, h(x))\, dx, \quad (42)$$

where $f$ and $h$ are arbitrary functions. For example, the above equation can be seen as a general form of $F_0(s)$ *i.e.*, Eq. (27). Replacing the integral variable $x$ by $h$, Eq. (42) can be rewritten as

$$F_1(s) = \frac{1}{x_2 - x_1} \int_{h(x_1)}^{h(x_2)} f(s, h) \frac{dx}{dh}\, dh. \quad (43)$$

We note that $dx/dh$ is a function of $h$ and it can be calculated from $1/(dh/dx)$. Here, we consider an inverse function of $h(x)$. If $h(x)$ is a one-to-one correspondence in the integral range from $x_1$ to $x_2$, its inverse function can be systematically obtained, which we express as $v(h) = x$. Derivative of the inverse function is expressed as

$$\frac{dv}{dh} = \frac{dx}{dh}. \quad (44)$$

In addition, a following relationship holds:

$$\frac{1}{x_2 - x_1}\int_{h(x_1)}^{h(x_2)} \frac{dv}{dh} dh = \frac{1}{x_2 - x_1}\int_{v(h(x_1))}^{v(h(x_2))} dv = 1. \quad (45)$$

Hence, $(dx/dh)/(x_2 - x_1)$ written in Eq. (43) can be seen as PDF (probability density function): $p(h)$. Then, Eq. (43) is rewritten as

$$F_2(s) = \int_{h(x_1)}^{h(x_2)} f(s,h)p(h)\, dh, \quad (46)$$

where we defined the integral containing PDF as $F_2(s)$ $(= F_1(s))$ to distinguish it from the original integral form (Eq. (42)). If $h(x)$ is not the one-to-one correspondence in the integral range in Eq. (42), its $F_2(s)$ can be expressed as

$$F_2(s) = \frac{1}{x_2 - x_1}\Bigg[(x_{m1} - x_1)\int_{h(x_1)}^{h(x_{m1})} f(s,h)p_{m1}(h)\, dh + (x_{m2} - x_{m1})\int_{h(x_{m1})}^{h(x_{m2})} f(s,h)p_{m2}(h)\, dh + \cdots + (x_2 - x_{mN})\int_{h(x_{mN})}^{h(x_2)} f(s,h)p_{mN+1}(h)\, dh\Bigg], \quad (47)$$

where $x_{mj}$ ($j = 1, 2, \ldots, N$) is a midpoint between the one-to-one correspondence domains and $p_{mj}(h)$ is PDF of $h$ within the corresponding integral range. When the number of the midpoints is $N$, the number of the integrals is $N + 1$. Since $p_{mj}(h)$ is zero outside the corresponding integral range, it can be simplified as

$$F_2(s) = \frac{1}{x_2 - x_1} \int_{h_{min}}^{h_{max}} f(s,h) \left( \sum_{j=1}^{N+1} p_{mj}(h)(x_{mj} - x_{m(j-1)}) \right) dh. \quad (48)$$

where $x_{m0}$ and $x_{mN+1}$ represent $x_1$ and $x_2$, respectively. $h_{min}$ and $h_{max}$ are the minimum and the maximum values of $h(x)$ in the domain from $x_1$ to $x_2$. By the way, we define $F_3(s)$ as follows:

$$F_3(s) = \int_0^1 f(s, b(u))\, du, \quad (49)$$

which has an analogous form compared with Eq. (42). In this stage, a concrete expression of $b(u)$ is not known, but it will be fabricated so that a following equation holds: $F_3(s) = F_2(s) = F_1(s)$. The concrete expression of $b(u)$ can be derived as follows. Eq. (48) can be written as the PDF integral form:

$$\int_0^1 f(s, b(u))\, du = \int_{h(0)}^{h(1)} f(s,h) p(h)\, dh. \quad (50)$$

Here, please recollect that $p(h)$ was made from $h(x)$ of the one-to-one correspondence domain through the three steps.

Step (a): Obtain the inverse function of $h(x)$ ($v(h) = x$).
Step (b): Differentiate $v(h)$ with respect to $h$.
Step (c): Divide the derivative of $v(h)$ by $x_2 - x_1$, which is equal to $p(h)$.

Using the same calculation steps (a)-(c), $p(h)$ can be obtained from $b(u)$. However, the concrete expression of $b(u)$ is not known in this stage. Instead, suppose $p(h)$ is known in this stage. In that case, it is considered that the concrete expression of $b(u)$ can be obtained by tracing the inverse route of steps (a)-(c).

Step (C): Multiply $p(h)$ by 1 ($= x_2 - x_1$).

Step (B): Integrate $p(h)$ with respect to $h$ (it corresponds to CDF of $p(h)$).

Step (A): Obtain the inverse function of CDF ($= b(u)$).

Introducing arbitrary examples, we calculated $F_3(s)$ by substituting $b(u)$ obtained from steps (C)-(A), and then we confirmed that $F_3(s)$ exactly matches $F_2(s)$ and $F_1(s)$. It indicates that if the integrated curve (or line segment) of "$h(x)$" takes the one-to-one correspondence and its gradient is always positive in the domain from $x_1$ to $x_2$, $h(x)$ can be also inversely calculated from $p(h)$ through steps (C)-(A). However, in the inverse calculation of $p(h)$, the steps (C), (B), and (A) are replaced as follows.

Step (C'): Multiply $p(h)$ by $(x_2 - x_1)$.

Step (B'): Integrate $(x_2 - x_1)p(h)$ with respect to $h$, which we mention it as the CDF in step (B').

Step (A'): Obtain the inverse function of the CDF in step (B'), which we express the obtained inverse function as $B(u)$. Finally, $h(x)$ is generated by using following equation, $h(x) = B(x - x_1)$.

After we derived the mathematical technique, we found that the concept of the integral technique in $F_2(s)$ is similar to that of the Lebesgue integration. In addition, we note that if one would like to calculate $F_1(s)$ with $f(s, h(x)) = \exp(-sh(x))$ but it is analytically impossible, we may recommend the researcher to use the integral form of $F_2(s)$ and the cumulant expansion. It may be one of the compromise schemes.

## Appendix C: Inverse transforms of specified $F_1(s)$ and $F_3(s)$ generate PDF

We have explained above that $F_1(s)$, $F_2(s)$, and $F_3(s)$ are equally connected. In this chapter, it is explained that the inverse Laplace and the inverse Fourier transforms of specified $F_1(s)$ and $F_3(s)$ generate PDF. In the integrand of $F_2(s)$, there is $p(h)$. Since $p(h)$ is PDF, it is completely 0 outside the integration range. Hence, $F_2(s)$ can be rewritten as

$$F_2(s) = \int_0^\infty f(s,h) p(h) \, dh, \qquad (51)$$

or

$$F_2(s) = \int_{-\infty}^\infty f(s,h) p(h) \, dh. \qquad (52)$$

If $f(s,h)$ is $\exp(-sh)$, Eq. (51) corresponds to the Laplace transform. Therefore, the inverse Laplace transform of $F_1(s)$ and $F_3(s)$ generates $p(h)$ as follows:

$$\begin{aligned}
&\frac{1}{2\pi i} \int_{\gamma-i\infty}^{\gamma+i\infty} \exp(sh) \left[ \frac{1}{x_2 - x_1} \int_{x_1}^{x_2} \exp(-sh(x)) \, dx \right] ds \\
&= \frac{1}{2\pi i} \int_{\gamma-i\infty}^{\gamma+i\infty} \exp(sh) \left[ \int_0^1 \exp(-sb(u)) \, du \right] ds = p(h), \qquad (53)
\end{aligned}$$

where $b(u)$ is the inverse function of CDF, i.e., the inverse function of cumulative PDF. The right-hand side of Eq. (53) can also be written as $p(h) H_1(h - h_{\min}) H_1(h_{\max} - h)$, where $h_{min}$ and $h_{max}$ are the minimum and the maximum values of $h(x)$ in the domain from $x_1$ to $x_2$ ($H_1$ is the Heaviside step function containing a property that $H_1(0) = 1$). It should be noted that $h(x) \geq 0$ in Eq. (53) ($x$ can takes both positive and negative values), because $p(h) \geq 0$ and the integration range of the variable $h$ in Eq. (51) is from 0 to $\infty$. This restriction originates from relationship between layouts of CDF and its inverse function. For example, when $h(x) = 1 + \sin x$, the inverse Laplace transform generates its PDF (arcsine distribution):

$$\frac{1}{2\pi i}\int_{\gamma-i\infty}^{\gamma+i\infty} \exp(sh)\left[\frac{1}{2\pi}\int_0^{2\pi} \exp(-s(1+\sin x))\,dx\right] ds$$
$$= \frac{1}{\pi\sqrt{1-(h-1)^2}} H_1(h) H_1(2-h). \qquad (54)$$

Since the analytical solution of the function in the square bracket above is $J_0(is)e^{-s}$ ($J_0$ is the Bessel function of the first kind of 0th order, see Eq. (4)), the inverse Laplace transform of $J_0(is)e^{-s}$ is the arcsine distribution above. Likewise,

$$\frac{1}{2\pi i}\int_{\gamma-i\infty}^{\gamma+i\infty} \exp(sh)\left[\frac{1}{2\pi m}\int_0^{2\pi m} \exp(-s(1+\sin x))\,dx\right] ds$$
$$= \frac{1}{\pi\sqrt{1-(h-1)^2}} H_1(h) H_1(2-h), \qquad (55)$$

where $m$ is an arbitrary natural number. Next, we consider the case where the integration range is not an integral multiple of $2\pi$. For example, when $m$ is *an infinitely large natural number* and $\alpha$ is an arbitrary real number ($\alpha \neq \infty$), we obtain the following equation (compare with Eq. (55)),

$$\frac{1}{2\pi i}\int_{\gamma-i\infty}^{\gamma+i\infty} \exp(sh)\left[\frac{1}{2\pi m+\alpha}\int_0^{2\pi m+\alpha} \exp(-s(1+\sin x))\,dx\right]ds$$
$$=\frac{1}{\pi\sqrt{1-(h-1)^2}}H_1(h)H_1(2-h). \qquad (56)$$

On the other hand, when *m* is *not an infinitely large natural number* and $\alpha$ is in a range $0<\alpha\leq\pi/2$, it is calculated as

$$\frac{1}{2\pi i}\int_{\gamma-i\infty}^{\gamma+i\infty} \exp(sh)\left[\frac{1}{2\pi m+\alpha}\int_0^{2\pi m+\alpha} \exp(-s(1+\sin x))\,dx\right]ds$$

$$=\frac{1}{2\pi m+\alpha}\left[\frac{2\pi m}{\pi\sqrt{1-(h-1)^2}}H_1(h)H_1(2-h)\right.$$

$$+\frac{2\alpha}{\{\pi-4\arcsin(\sqrt{(1-\sin\alpha)/2})\}\sqrt{1-(h-1)^2}}H_1(h-1)H_1(1+\sin\alpha$$

$$\left.-h)\right]. \qquad (57)$$

In the calculation above, we made use of Eq. (48) and an equation below:

$$\frac{1}{2\pi i}\int_{\gamma-i\infty}^{\gamma+i\infty} \exp(sh)\left[\frac{1}{\alpha}\int_0^{\alpha} \exp(-s(1+\sin x))\,dx\right]ds$$
$$=\frac{1}{\pi\sqrt{1-(h-1)^2}}H_1(h-1)H_1(1+\sin\alpha-h)$$
$$/\int_1^{1+\sin\alpha}\frac{1}{\pi\sqrt{1-(h'-1)^2}}dh'. \qquad (58)$$

As far as we know, an analytical solution of the integral on *x* in Eq. (57) is not known (it cannot be analytically solved), however, the inverse Laplace transform of "the integral on *x*" can be obtained from our mathematical technique as shown in the right-hand side of Eq. (57). It is an interesting point of the technique.

This kind of the inverse calculation can be also performed in the case of the inverse Fourier transform. The inverse Fourier transform of $F_1(s)$ and $F_3(s)$ generates $p(h)$ as follows:

$$\frac{1}{2\pi}\int_{-\infty}^{\infty} \exp(ish) \left[\frac{1}{x_2-x_1}\int_{x_1}^{x_2} \exp(-ish(x))\,dx\right] ds$$
$$= \frac{1}{2\pi}\int_{-\infty}^{\infty} \exp(ish) \left[\int_0^1 \exp(-isb(u))\,du\right] ds = p(h). \quad (59)$$

The right-hand side of Eq. (59) can also be written as $p(h)H_1(h-h_{min})H_1(h_{max}-h)$, where $h_{min}$ and $h_{max}$ are the minimum and the maximum values of $h(x)$ in the domain from $x_1$ to $x_2$. We note that $x$ can take both positive and negative values. Since the integration range of the Fourier transform of $p(h)$ is from $-\infty$ to $\infty$, $h$ can take both positive and negative values.

In studies of liquids [9], colloidal dispersions [10], small angle scatterings [11][12], the Fourier transform for a radial distribution function (the spherical symmetric Fourier transform) is often used. Using its forward and inverse transforms, a following mathematical relationship can be also written:

$$\frac{2}{\pi}\int_0^{\infty} \frac{s\sin(sh)}{h}\left[\frac{1}{x_2-x_1}\int_{x_1}^{x_2} \frac{h(x)\sin(sh(x))}{s}\,dx\right] ds$$
$$= \frac{2}{\pi}\int_0^{\infty} \frac{s\sin(sh)}{h}\left[\int_0^1 \frac{b(u)\sin(sb(u))}{s}\,du\right] ds = p(h). \quad (60)$$

The right-hand side of Eq. (60) can also be written as $p(h)H_1(h-h_{min})H_1(h_{max}-h)$, where $h_{min}$ and $h_{max}$ are the minimum and the maximum values of $h(x)$ in the domain from $x_1$ to $x_2$. It should be noted that $x$ can take both positive and negative values and $h(x) \geq 0$ in Eq. (60). The restriction of $h(x) \geq 0$ originates from $p(h) \geq 0$ and the integration range of the variable $h$ in the spherical symmetric Fourier transform.

Generally, an integral transform (forward version) can be expressed as

$$F(s) = \int_{r_1}^{r_2} K_{\text{for}}(s,h)p(h)\,dh, \qquad (61)$$

where $K_{\text{for}}$ represents the integral kernel of the forward transform. $r_1$ and $r_2$ follow an integral range of the forward transform. In addition, the corresponding inverse transform is written as

$$p(h) = \int_{r_3}^{r_4} K_{\text{inv}}(s,h)F(s)\,ds, \qquad (62)$$

where $K_{\text{inv}}$ represents the integral kernel of the inverse transform. $r_3$ and $r_4$ follow an integral range of the inverse transform. For example, Hankel, Mellin, Hartley, Kontorovich-Lebedev, and Laguerre transforms are also categorized as the integral transform. Applying the findings obtained from the present study, the following relationship can be written:

$$\int_{r_3}^{r_4} K_{\text{inv}}(s,h)\left[\frac{1}{x_2-x_1}\int_{x_1}^{x_2} K_{\text{for}}(s,h(x))dx\right]ds$$
$$= \int_{r_3}^{r_4} K_{\text{inv}}(s,h)\left[\int_0^1 K_{\text{for}}(s,b(u))du\right]ds = p(h) \qquad (63)$$

It should be noted that the value of $h(x)$ and the domain from $x_1$ to $x_2$ follows each rule of the integral transform and the layout. $p(h)$ is PDF of the $h(x)$ in the domain from $x_1$ to $x_2$. The right-hand side of Eq. (63) can also be written as $p(h)H_1(h - h_{\min})H_1(h_{\max} - h)$, where $h_{\min}$ and $h_{\max}$ are the minimum and the maximum values of $h(x)$ in the domain from $x_1$ to $x_2$. In addition, the forward transform of $p(h)$ with the integral kernel $K_{\text{inv}}$ can be written as

$$\int_{r_1}^{r_2} K_{\text{for}}(s,h)p(h)\,dh = \frac{1}{x_2 - x_1}\int_{x_1}^{x_2} K_{\text{for}}\big(s,h(x)\big)dx = \int_0^1 K_{\text{for}}\big(s,b(u)\big)du. \qquad (64)$$

It expresses a general form of the following relationship $F_2(s) = F_1(s) = F_3(s)$.

**References**

[1]   L. Ferrero, M. Pissavini, S. Marguerie, L. Zastrow, Efficiency of a continuous height distribution model of sunscreen film geometry to predict a realistic sun protection factor, J. Cosmet. Sci. 54 (2003) 463–481.

[2]   L. Ferrero, M. Pissavini, O. Doucet, How a calculated model of sunscreen film geometry can explain in vitro and in vivo SPF variation, Photochem. Photobiol. Sci. 9 (2010) 540–551. https://doi.org/10.1039/b9pp00183b.

[3]   X. Zhang, H. Ejima, N. Yoshie, Periodic nanopatterns from polymer blends via directional solidification and subsequent epitaxial crystallization, Polym. J. 47 (2015) 498–504. https://doi.org/10.1038/pj.2015.26.

[4]   S. Kodama, X. Zhang, N. Yoshie, Formation of nanostructured thin films of immiscible polymer blends by directional crystallization onto a crystallizable organic solvent, Colloid Polym. Sci. 293 (2015) 2165–2169. https://doi.org/10.1007/s00396-015-3593-9.

[5]   I. Kato, T. Tanaka, K. Okoshi, Dilated Smectic Liquid Crystal of Polystyrene- block -polysilane- block -polystyrene Copolymer Synthesized by Atom Transfer Radical Polymerization , Chem. Lett. 49 (2020) 347–349. https://doi.org/10.1246/cl.200071.

[6]   J.S. Pedersen, Small-angle scattering from precipitates: Analysis by use of a

polydisperse hard-sphere model, Phys. Rev. B. 47 (1993) 657–665. https://doi.org/10.1103/PhysRevB.47.657.

[7] P.J. Flory, Molecular Size Distribution in Linear Condensation Polymers, J. Am. Chem. Soc. 58 (1936) 1877–1885. https://doi.org/10.1021/ja01301a016.

[8] P. Bryk, M. Bryk, Effective interactions in polydisperse colloidal suspensions investigated using Ornstein-Zernike integral equations, J. Colloid Interface Sci. 338 (2009) 92–98. https://doi.org/10.1016/j.jcis.2009.05.078.

[9] S. Woelki, H.H. Kohler, H. Krienke, A singlet-RISM theory for solid/liquid interfaces part I: Uncharged walls, J. Phys. Chem. B. 111 (2007) 13386–13397. https://doi.org/10.1021/jp068998t.

[10] K. Amano, T. Hayashi, K. Hashimoto, N. Nishi, T. Sakka, Potential of mean force between spherical particles in an ionic liquid and its decomposition into energetic and entropic components: An analysis using an integral equation theory, J. Mol. Liq. 257 (2018) 121–131. https://doi.org/10.1016/j.molliq.2018.02.089.

[11] T. Fukasawa, T. Sato, Versatile application of indirect Fourier transformation to structure factor analysis: From X-ray diffraction of molecular liquids to small angle scattering of protein solutions, Phys. Chem. Chem. Phys. 13 (2011) 3187–3196. https://doi.org/10.1039/c0cp01679a.

[12] K. Amano, R. Sawazumi, H. Imamura, T. Sumi, K. Hashimoto, K. Fukami, H. Kitaoka, N. Nishi, T. Sakka, An improved model-potential-free analysis of the structure factor obtained from a small-angle scattering: Acquisitions of the pair distribution function and the pair potential, Chem. Lett. 49 (2020) 1017–1021. https://doi.org/10.1246/CL.200292.